\documentclass[doublecol]{epl2}

\usepackage{amsmath}
\usepackage{amstext}
\usepackage{amssymb}
\usepackage{latexsym}

\newcommand{\beq}{\begin{equation}}
\newcommand{\eeq}{\end{equation}}
\newcommand{\Ham}{\mathcal H}

\title{Photon and polariton fluctuations in arrays of QED-cavities}
\shorttitle{Fluctuations in cavity arrays}

\author{Davide Rossini\inst{1} \and Rosario Fazio\inst{1,2} \and Giuseppe Santoro\inst{1,3,4}}

\institute{
  \inst{1} International School for Advanced Studies (SISSA),
           Via Beirut 2-4, I-34014 Trieste, Italy

  \inst{2} NEST-CNR-INFM \& Scuola Normale Superiore,
           Piazza dei Cavalieri 7, I-56126 Pisa, Italy

  \inst{3} CNR-INFM Democritos National Simulation Center, 
           Via Beirut 2-4, I-34014 Trieste, Italy

  \inst{4} International Centre for Theoretical Physics (ICTP),
           P.O.Box 586, I-34014 Trieste, Italy}


\abstract{
We propose to detect the Mott insulator-superfluid quantum phase 
transition in an array of coupled cavities by studying the polariton and
photon fluctuations in a block of linear dimension $M$ (in units of the
lattice constant of the array). We explicitly show this for a one-dimensional
array; the analysis can be however extended to higher dimensions. 
In the Mott phase polariton fluctuations are independent of the block size. 
In the superfluid phase they grow logarithmically with $M$, the prefactor 
being related to the compressibility of the system. In the case of photon 
fluctuations, the critical behaviour is encoded in the subleading scaling with 
the block dimension, while the leading behaviour is linear in $M$ and non-critical. 
Our results have been obtained by means of the density matrix renormalization 
group numerical algorithm.
}

\begin{document}

\maketitle

The recent suggestion~\cite{hartmann06,greentree06,angelakis06}
to realize Mott and superfluid phases in arrays of coupled QED-cavities 
has stimulated a flurry of activity on strongly interacting photonic 
systems~\cite{neilna07,rossini07,paternostro07,hartmann07,hartmann07B,cho08,kay08,makin07,
aichhorn07}.
In the presence of randomness a glassy phase of polaritons was 
shown to appear in the phase diagram~\cite{rossini07}.
Coupled cavities can be engineered to behave as quantum simulators for a
variety of interacting spin models~\cite{hartmann07,cho08,kay08};
they can support soliton excitations~\cite{slusher03,paternostro07}.
A different realization of the strongly interacting regime with photons
in an optical guide was proposed in~\cite{chang07}, while polariton blockade effects
were studied in a resonantly excited photonic quantum dot~\cite{verger06}.
As candidates for simulating strongly interacting models, coupled cavities present 
new characteristics as compared to other systems, like optical
lattices~\cite{lewenstein06} or Josephson junction arrays~\cite{fazio01}. 
Most notably, it is possible to access their local properties. 

In this paper we would like to exploit the local addressability of cavity arrays to 
propose a method in order to probe the critical behaviour of the Mott insulator-superfluid 
transition. Our suggestion is based on the study of the fluctuations in the number 
of photons and polaritons. To our knowledge, this issue was not addressed so far 
in the literature. While it is well known how to distinguish a regime of photon 
blockade, how to detect the transition itself is, to a large extent, an unexplored 
problem. We study in details a one-dimensional array. Our conclusions, however, can 
be easily extended to higher dimensions.

{\em The Model - } 
We suppose that inside each cavity a single two-level atom interacts with photons 
via a Jaynes-Cummings Hamiltonian~\cite{jcreview93}. While this situation is simpler 
to simulate, we do not expect, for the purposes of our work, any qualitative change 
if other atomic level structures~\cite{hartmann06} are considered. 
The Hamiltonian for an array of $L$ coupled cavities is then given by
\beq
   \Ham = \sum_{i=1}^L \left[ \epsilon \, \sigma_i^+ \sigma_i^-
   + \omega \, a^\dagger_i a_i
   + \beta \big( \sigma^+_i a_i + h.c.\big) \right] 
   - t \sum_{\left\langle i,j \right\rangle} a^\dagger_i a_j
   \label{eq:fullham}
\eeq
where $\epsilon$ denotes the transition energy between the two atomic
levels, $\omega$ is the resonance frequency of the cavity, $\beta$
is the atom-field coupling constant ($\epsilon, \, \omega, \, \beta > 0$),
and $t$ is the inter-cavity photon hopping, that is assumed to be
constant for all nearest neighbours and zero otherwise.
The atomic and photonic raising/lowering operators are denoted as $\sigma_{i}^{\pm}$,
and $\{ a^\dagger_i , a_i \}$ respectively, the subscript $i$ indicating the 
lattice site. The total number of atomic plus photonic excitations
(i.e., the number operator for polaritons) on the $i$-th site is given by
$n_i^{\rm pol} = n_i^{\rm ph} + \sigma^+_i \sigma^-_i$, where
$n_i^{\rm ph} = a^\dagger_i a_i$ is the photon number operator.
Our analysis is restricted to the case of zero relative detuning
$\Delta \equiv \omega - \epsilon$; we also work in the canonical ensemble
with a fixed polariton density $\rho \equiv N/L = 1$,
$N$ being the total number of polaritons in $L$ cavities.

The equilibrium phase diagram associated to the model in Eq.~\eqref{eq:fullham} is 
characterized by two distinct phases, with polariton Mott Insulating (MI) regions 
surrounded by the Superfluid (SF) phase. In the MI phase polaritons are localized on 
each site due to the photon blockade~\cite{photonblockade}, and there is a gap in the 
spectrum. A finite hopping renormalizes this gap, which eventually vanishes at a critical 
value of $t$. For large hoppings excitations are delocalized and the system enters 
the SF phase. 

We propose to detect the MI-SF transition by analyzing the fluctuations of the
occupation number inside a block composed by a subset of $M$ adjacent cavities.
Since the number operator is the canonically conjugated variable with respect
to the phase of the whole many-body wavefunction, we expect that its
fluctuations are strongly suppressed in the incoherent MI regime, and,
by contrast, greatly enhanced in the coherent SF phase.
The dispersion of particle number on a given subsystem with $M$ sites
is quantified by the variance of the corresponding probability distribution:
\beq
   \delta n^2_{\alpha} (M) = \Big\langle \big( \sum_{i \in M} n_i^{\alpha} \big)^2 \Big\rangle
   - \Big\langle \sum_{i \in M} n_i^{\alpha} \Big\rangle^2 
\eeq
where $\alpha={\rm pol}/{\rm ph}$ stands for polariton/photon fluctuations,
and $\langle \cdot \rangle$ denotes an average on the system ground state.
This can be obtained from the two-point correlation functions of the related
number operator:
$C_{ij}^{\alpha} \equiv \langle n_i^{\alpha} n_j^{\alpha} \rangle -
\langle n_i^{\alpha} \rangle \langle n_j^{\alpha} \rangle$, such that
$\delta n^2_{\alpha} (M) = \sum_{(i,j) \in M} C_{ij}^{\alpha}$.

In the present work we suppose that the system is in its ground state; we neglect 
decoherence, and assume that spontaneous photon emission and cavity loss characteristic 
times are much longer than the timescale over which the array can reach the 
ground state. We are aware that this is a strong assumption that may not 
be fulfilled in the experiments. We do not try to argue on this very important 
issue in the present work; rather we are interested in describing a method that is
suitable to detect the various (equilibrium and non-equilibrium) many-body phases
of an array of cavities. We describe our proposal by using the (equilibrium) MI-SF 
transition; the essential features of the method can be as well applied to 
the non-equilibrium case. We will come back to this point in the concluding section. 

All the numerical data presented below have been obtained by means of the 
Density Matrix Renormalization Group (DMRG) algorithm with open boundary conditions.

%%%%%%%%%%%%%%%%%%%%%%%%%%%%%
\begin{figure}[!t]
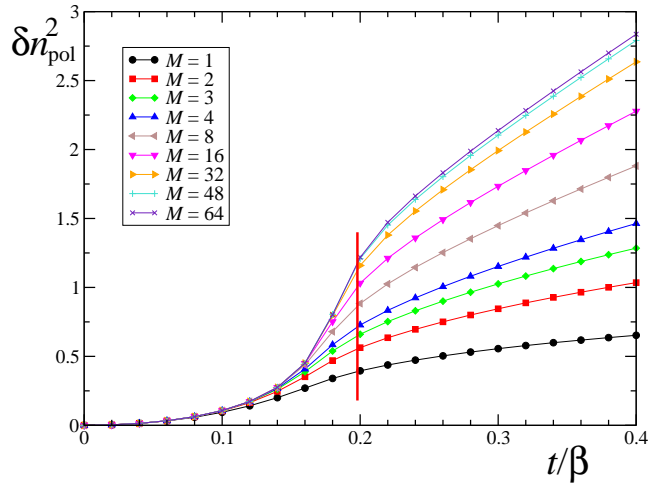

    \onefigure[scale=0.34]{Pol_tvar}
    \caption{(Colour on-line) Variance $\delta n^2_{\rm pol} (M)$ of the probability
      distribution for the polaritonic occupation number inside a block of $M$ cavities
      as a function of the inter-cavity photon hopping $t$, for fixed values of $M$.
      With our choice of parameters, the MI-SF transition point is at
      $t^*/\beta \approx 0.198$ (red solid line).
      DMRG simulations have been performed for a system of $L=128$ cavities, keeping
      a maximum of $n_{\rm max} = 10$ polaritons per site and a number $m=150$
      of states; blocks are formed starting from the center of the chain.}
      \label{fig:Pol_tvar}
\end{figure}
%%%%%%%%%%%%%%%%%%%%%%%%%%%%%

{\em Polariton fluctuations - } We first study fluctuations in the polariton number.
The number of polaritons inside a given subsystem can be experimentally measured by
instantaneously switching off the effective polaritonic hopping (this can be
achieved by changing the detuning $\Delta$, such to increase the relative
strength of the atom-field coupling with respect to the photon hopping).
In this way, admitting radiative losses on long time scales, a quantum-jump picture
immediately suggests that the polariton number is exactly given by the number
of photons emitted from the selected region~\cite{plenio98}.

In Fig.~\ref{fig:Pol_tvar} we show numerical data displaying
$\delta n^2_{\rm pol} (M)$ as a function of the photon hopping $t$
in a system with $L=128$ cavities; various curves correspond to
different sizes $M$ of the block inside the system.
The MI-SF transition point has been located numerically with high
accuracy at $t^*/ \beta \approx 0.198$~\cite{rossini07}.
On-site fluctuations cannot reveal critical features at the phase transition,
because they correspond to a local property; whenever $t \neq 0$,
the solid curve with circles corresponding to $M=1$ has always non-zero values.
Therefore, an unambiguous characterization of the phase boundary is impossible
in this context.
This was already pointed out in Ref.~\cite{capo07} for the on-site boson
number fluctuations in the Bose-Hubbard model, where it has been
shown that the number probability distribution evolves from a Poissonian,
in the noninteracting gas, to a sharply peaked distribution, in the insulator.
We obtained very similar distribution probabilities for the on-site polariton
number in our system, that are characterized by sub-Poissonian statistics.

\begin{figure}[!t]
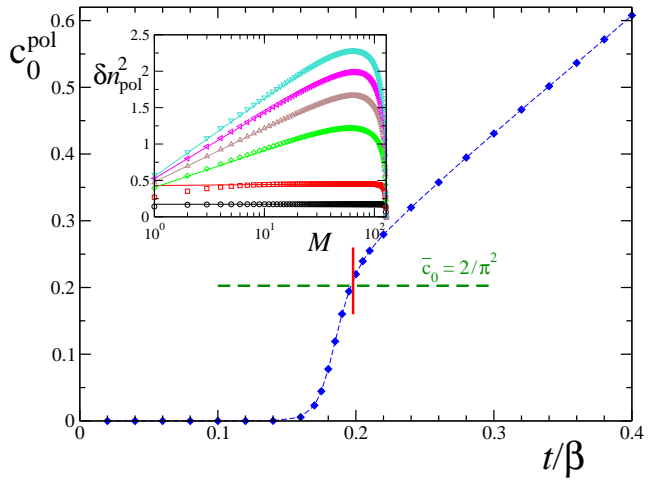

    \onefigure[scale=0.34]{Pol_C0_cntr}
    \caption{(Colour on-line) Best fitting parameter $c^{\rm{pol}}_0$ of Eq.~\eqref{eq:PolFit_log}
      for the variance $\delta n^2_{\rm pol} (M)$ as a function of the hopping $t$.
      The horizontal dashed line indicates the estimate of the logarithmic
      prefactor at the transition point: $\bar{c}_0 \sim 2 / \pi^2$.
      In the inset we plot $\delta n^2_{\rm pol}$ as a function of $M$ (symbols)
      for some fixed values of $t$: from bottom to top
      $t=0.12, \, 0.16, \, 0.2, \, 0.24, \, 0.28, \, 0.32$;
      continuous lines are logarithmic fits of the corresponding data.
      In numerical fits we dropped the first and the latter 10 points
      (i.e., we kept values for $M \in [11, L/2-10]$).}
%      The same DMRG data of Fig.~\ref{fig:Pol_tvar} have been used.}
    \label{fig:Pol_C0}
\end{figure}

From Fig.~\ref{fig:Pol_tvar} we notice that, by increasing $M$, in the MI phase
fluctuations tend to be independent on the block size, while in the SF phase
the dependence on $M$ is evident.
In particular, we found that $\delta n^2_{\rm pol} (M)$ as a function of $M$
saturates to a finite constant value in the MI phase, while it diverges
logarithmically in the SF phase (see inset of Fig.~\ref{fig:Pol_C0}, where we
plotted the same curves of Fig.~\ref{fig:Pol_tvar}, but fixing $t$ and varying $M$).
The logarithm divergence is directly related to the fact that density correlation
functions in the superfluid phase have a power-law decaying uniform term~\cite{haldane81}:
\beq
C_{r}^{\rm pol} =
\langle n_{i+r}^{\rm pol} n_i^{\rm pol} \rangle -
\langle n_{i+r}^{\rm pol} \rangle \langle n_i^{\rm pol} \rangle \sim \frac{2}{K}
(2 \pi \rho r)^{-2} + \cdots \, .
\eeq
In the previous expression $\rho$ is the particle (in this case the polariton) 
density and $K$ is the so called Luttinger parameter~\cite{giamarchi92}, that is
proportional to the square root of the compressibility of the system.
Indeed, from the definition of $\delta n^2_{\rm pol} (M)$, one immediately concludes
that, in the SF phase 
\beq
\delta n^2_{\rm pol}(M) \sim \frac{1}{K \pi^2 \rho^2} \ln M \, .
\label{LL}
\eeq
At the SF-MI transition the coefficient $K$ at integer densities is known to be equal 
to $K_c = 1/2$~\cite{giamarchi92}.
Therefore, for a polariton density $\rho=1$, the logarithmic prefactor suddenly
jumps from a value $\bar{c}_0 = 2/\pi^2$ just inside the SF phase, to zero
in the MI, with a characteristic Kosterlitz-Thouless behaviour.

A quantitative analysis of the crossover between the two different behaviours has been 
performed by fitting numerical data according to:
\beq
   \delta n^2_{\rm pol} (M) =
   c^{\rm{pol}}_0 \ln \left[ \frac{L}{\pi} \sin \left( \frac{\pi}{L} M \right) \right] + 
   A^{\rm{pol}} \, ,
   \label{eq:PolFit_log}
\eeq
with $A^{\rm{pol}}$ and $c^{\rm{pol}}_0$ as fitting parameters. The constant term  $A^{\rm{pol}}$
is not important for our purposes, therefore we concentrate on the logarithmic term. 
In the main panel of Fig.~\ref{fig:Pol_C0} we plot the logarithmic prefactor $c^{\rm{pol}}_0$
as a function of $t$: the phase transition point in this situation can be quite
clearly identified. In 1D, in particular, a precise criterion for values of
$c^{\rm{pol}}_0 \ge 2/\pi^2$ in the superfluid phase, clearly identifies the transition point.

%%%%%%%%%%%%%%%%%%%%%%%%%%%%%
%%%%%%%%%%%%%%%%%%%%%%%%%%%%%

{\em Photon fluctuations - } 
We now concentrate on the photon number fluctuations, a quantity which is more directly
measured in quantum optical experiments, where the system is typically
in a non-equilibrium state between photon/cavity decay and external pumping.
The state of the polariton field is usually retrieved by detecting and
characterizing the light emission~\cite{carusotto05,verger07}.

\begin{figure}[!t]
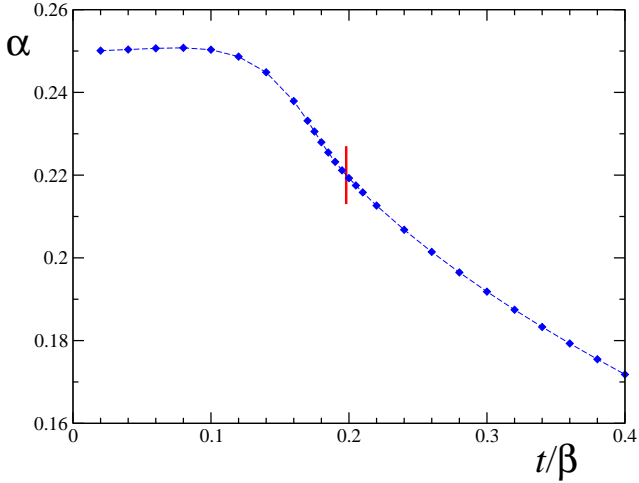

    \onefigure[scale=0.34]{Phot_alpha}
    \caption{(Colour on-line) Coefficient of the linear term in the photon fluctuations,
    see Eq.~\eqref{eq:PhotFit_log}, as a function of the photon hopping $t$ between the 
    cavities. As it is evident from the plot, $\alpha$ carries no information 
    about the critical behaviour (as a reference, we indicated with a red bar the 
    location of the critical point).}
    \label{alphascal}
\end{figure}

\begin{figure}[!t]
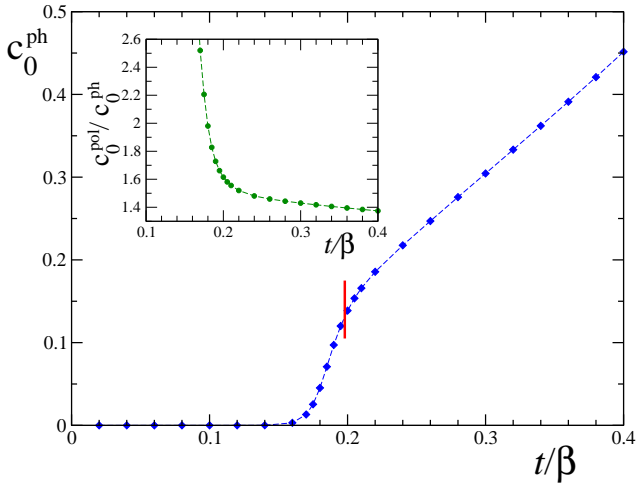

    \onefigure[scale=0.34]{Phot_C0_cntr}
    \caption{(Colour on-line) Prefactor $c^{\rm ph}_0$ of the logarithmic term
    in Eq.~\eqref{eq:PhotFit_log} as a function of the hopping.
    The behaviour is very similar to the corresponding prefactor for the polariton
    fluctuations. The ratio between $c^{\rm{pol}}_0$ and 
    $c^{\rm ph}_0$, shown in the inset, is non-universal.}
    \label{c0tildescal}
\end{figure}

In this case the situation 
is quite different as compared to the case of polariton fluctuations.  
Even at zero hopping the on-site photon number is fluctuating.
For example, in the case of zero relative detuning ($\Delta = 0$) and deep
in the MI regime, we have $\langle n_i^{\rm ph} \rangle \approx 1/2$ and
$\langle (n_i^{\rm ph} )^2 \rangle \approx 1/2$. Therefore, even for a perfect insulator, 
the variance $\delta n^2_{\rm ph} (M)$ of the photon number distribution inside a block
of length $M$ is proportional to the block size: $\delta n^2_{\rm ph} (M) \approx M/4$.
The onset of the superfluid behaviour can then be sought in the deviations from this 
linear growth. These deviations are due to the raising of correlations
between distant sites on increasing the hopping strength. Therefore, in the scaling 
ansatz for the photon fluctuations one should also include a term that is linear
in the block dimension, i.e.
\beq
   \delta n^2_{\rm ph} (M) = \alpha M +
   c^{\rm{ph}}_0 \ln \left[ \frac{L}{\pi} \sin \left( \frac{\pi}{L} M \right) \right] + 
   A^{\rm{ph}} \, .
   \label{eq:PhotFit_log}
\eeq
Let us analyze the properties of the first two terms in the r.h.s. of 
Eq.~\eqref{eq:PhotFit_log}.
The behaviour of $\alpha$ as a function of the hopping is shown in Fig.\ref{alphascal}. 
Starting from the value $\alpha=1/4$ at $t=0$, the coefficient of the linear term  
decreases on increasing the hopping. This behaviour signals the fact that correlations
between photon fluctuations at different locations start to develop.
%%%%%%%%%%%%%%%%%%%%%%%%%%%%%
%%%%%%%%%%%%%%%%%%%%%%%%%%%%%
Although at $t\sim 0.2$ there is an indication of the change in the ground state 
properties, the coefficient $\alpha$ seems not to be an appropriate indicator of the 
location of the critical point. As for the case of polariton 
fluctuations, also in the case of photons one should look at the coefficient of the 
logarithmic term. This is shown in  Fig.~\ref{c0tildescal}: here the critical behaviour is 
easily identified. The value of the fluctuation at the transition is not universal 
as in the polariton case; the reason is that photons are not the genuine excitations 
of the system, therefore the considerations leading to Eq.~\eqref{LL} with $K_{c}=1/2$ 
do not apply.
%%%%%%%%%%%%%%%%%%%%%%%%%%%%%
%%%%%%%%%%%%%%%%%%%%%%%%%%%%%
The ratio between the coefficients of polariton and photon fluctuations is shown 
in the inset of Fig.~\ref{c0tildescal}. A marked change at the transition can be 
noticed. In our opinion, however, this is not obviously related to the critical 
point; rather we expect that the ratio is a non-universal feature depending on the 
details of the model and on the value of the couplings. 

Indication of the critical behaviour from photon correlations may also be 
obtained without resorting to a scaling analysis as a function of the block 
size $M$. This can be achieved by studying the second derivative of photon 
number fluctuations with respect to the block size, evaluated for blocks 
of half the length of the system size, that is 
$\partial^2_M \big[ \delta n^2_{\rm ph} (M) \big] \big\vert_{M = L/2}$. Although this 
approach might be slightly less accurate, it can give an interesting insight into 
the critical region.

\begin{figure}[!t]
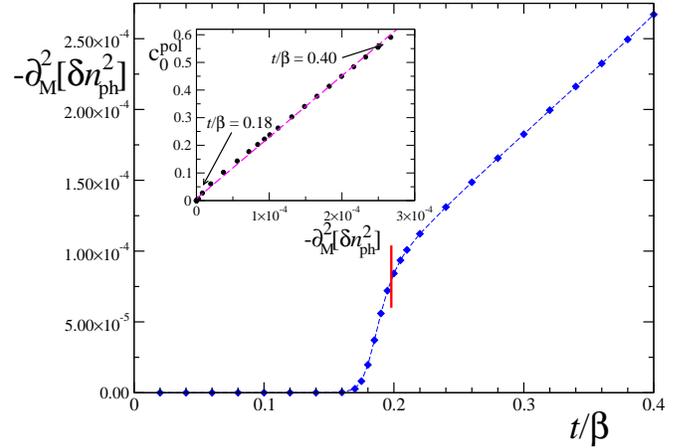

    \onefigure[scale=0.32]{Phot_cnt_2Der}
    \caption{(Colour on-line) Second derivative of photon number fluctuations with
      respect to the block length, evaluated at half the length of the system size,
      as a function of the hopping $t$.
      Inset: logarithmic prefactor of polariton number fluctuations, as
      a function of the second derivative of photon number fluctuations
      (data are the same of the main panel and of Fig.~\ref{fig:Pol_C0}).}
    \label{fig:Phot_2Der}
\end{figure}

We first consider $\partial^2_M \big[ \delta n^2_{\rm ph} (M) 
\big] \big\vert_{M=L/2}$ as a function of the hopping parameter $t$. 
Numerical data in Fig.~\ref{fig:Phot_2Der} show that, like for the polaritonic
number fluctuations, also this quantity can be used to characterize the MI-SF
transition for finite sizes.
In addition, we can identify a behaviour that is very similar to the one
of the logarithmic prefactor $c^{\rm{pol}}_0$ of Eq.~\eqref{eq:PolFit_log} for the
polariton number fluctuations.
Namely, the correlation between these two quantities is nearly perfect:
apart from a proportionality factor, which depends on the system size,
the two curves in the main panels of Figs.~\ref{fig:Pol_C0}-\ref{fig:Phot_2Der} 
are exactly the same.
This is shown in the inset of Fig.~\ref{fig:Phot_2Der} where, for a given value
of the photon hopping $t$, we plot the corresponding values of $c_0^{\rm pol}$ and
of $\partial^2_M \big[ \delta n^2_{\rm ph} (M) \big] \big\vert_{M=L/2}$ in the
two axes, thus displaying a perfect linear correlation.

%%%%%%%%%%%%%%%%%%%%%%%%%%%%%
%%%%%%%%%%%%%%%%%%%%%%%%%%%%%

\begin{figure}[!t]
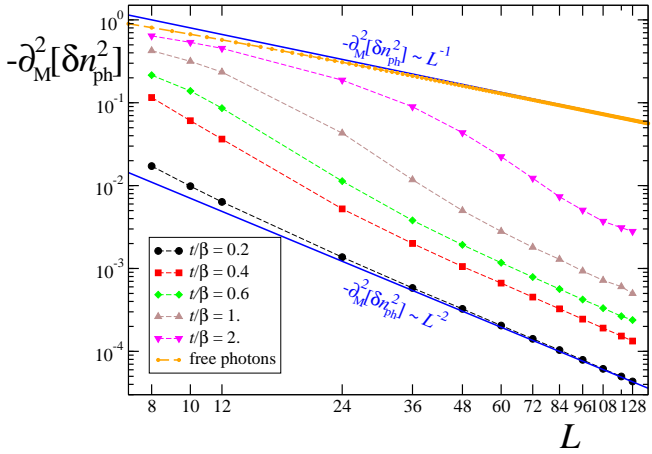

    \onefigure[scale=0.32]{Phot_2Der_scal}
    \caption{(Colour on-line) Second derivative of photon number fluctuations with
      respect to the block size $M$ evaluated at $M=L/2$, as a function
      of the system size $L$.
      The two straight blue lines denote power-law behaviours $L^{-1}, \,  L^{-2}$.
      DMRG parameters are the same as in Figs.~\ref{fig:Pol_tvar} and~\ref{fig:Pol_C0}.}
    \label{fig:Phot_2Der_scal}
\end{figure}

The dependence of the photon fluctuations on the size is however more complicated than
that of $\delta n^2_{\rm pol}$. 
In Fig.~\ref{fig:Phot_2Der_scal} we plot
$\partial^2_M \big[ \delta n^2_{\rm ph} (M) \big] \big\vert_{M=L/2}$ 
(its absolute value, actually, since it is always negative) as a function of the
system size $L$, for several fixed photon hoppings $t$.
While the logarithmic prefactors $c^{{\rm pol}}_0$ and $c^{{\rm ph}}_0$ are independent of 
the system size, this is not the case for the second derivative of photon 
number fluctuations, which asymptotically drops to zero as a power-law with $L$.
More specifically, in the free photon limit ($t \to +\infty$) it decays
as $L^{-1}$, while for finite values of the photon hopping and a sufficiently large
system size, there is a crossover to a $L^{-2}$ behaviour.
This follows from two qualitatively different decays of the photon number
correlation function 
$C_{r}^{\rm ph} = \langle n_{i+r}^{\rm ph} n_i^{\rm ph} \rangle - 
\langle n_{i+r}^{\rm ph} \rangle \langle n_i^{\rm ph} \rangle$:
apart from open-boundary effects,
in the first regime $C_r^{\rm ph} \approx 1/L$, while in the second case
$C_r^{\rm ph} \approx 1/r^2$.
We carried out a numerical analysis of the photon correlations $C_r^{\rm ph}$
in our QED-cavity array system, and explicitly found these two distinct behaviours.

%%%%%%%%%%%%%%%%%%%%%%%%%%%%%
%%%%%%%%%%%%%%%%%%%%%%%%%%%%%

It has also been possible to derive an analytic estimate of the crossover scale
by exploiting a mapping of the cavity array with a Bose-Hubbard (BH) model,
and studying its corresponding boson correlators $C_{ij}^{\rm BH}$.
It should be kept in mind that the formal analogy between $C_{ij}^{\rm ph}$
and $C_{ij}^{\rm BH}$ has to be considered only at a qualitative level,
since the equivalence of the two models becomes exact only in the limit
of a large number of atoms per cavity~\cite{hartmann06}.

The BH Hamiltonian is defined by
\beq
   \Ham = - J \sum_j ( b^\dagger_{j+1} b_j + h.c.) %b^\dagger_j b_{j+1} )
   + \frac{U}{2} \sum_{j=1}^L b^\dagger_j b^\dagger_j b_j b_j \, ,
   \label{eq:BHmodel}
\eeq
where $\{ b_j^\dagger, b_j \}$ are the boson creation/annihilation operators, 
$n_i = b^\dagger_i b_i$ the boson number operator, and with $C_{ij}^{\rm BH}$
we indicate the corresponding correlation functions.
When the depletion of the condensate is not too great ($J \gg U$),
we can employ the Bogoliubov approximation, which consists in replacing the
boson creation and annihilation operators at a given site $j$
by a c-number $z_i \in \mathbb{C}$ plus a fluctuation operator $\beta_j$.
Only terms at most of the second order in $\beta_j$ are considered when
diagonalizing the Hamiltonian~\cite{oosten01,rey03,burnett02}.
Within this framework and for periodic boundary conditions, we
locate the crossover scale between the $L^{-1}$ and $L^{-2}$ behaviour by
\beq
U/J \lesssim \frac{2 \pi^2}{n_0L^2} \,,
\eeq
with $n_0$ being the density of the boson condensate.
Whenever the inequality is valid, the spectrum of the Bogoliubov quasi-particles
at small $k$ is quadratic, and correlations behave as for free bosons:
$C_r^{\rm BH} \approx 1/L$.
Conversely, if it does not hold, the spectrum becomes linear at small $k$,
and the free-boson limit fails; in this regime we recover the decay
$C_r^{\rm BH} \approx 1/r^2$ for large $r$,
typical of the polariton number correlations~\cite{haldane81}.

\begin{figure}[!t]
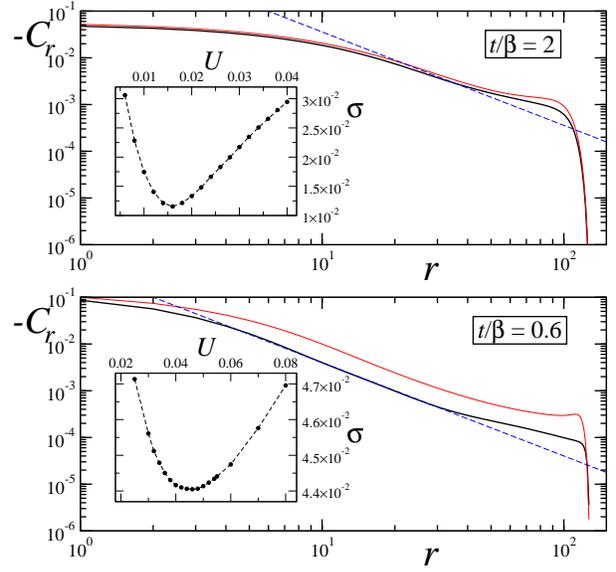

    \onefigure[scale=0.36]{Phot-BH_2}
    \caption{(Colour on-line) Number fluctuations of photons $C_r^{\rm ph}$ in an array
      of cavities (black curves - DMRG data) and of bosons $C_r^{\rm BH}$ in
      BH model (red curves - Bogoliubov ansatz);
      dashed blue curves display a behaviour $C_r \sim r^{-2}$.
      The two panels correspond to different values of the hopping,
      that is assumed to be equal in the two models $t = J$.
      In the insets we plot the deviation $\sigma$ of correlations for the two models,
      as a function of $U$; the red curve in each panel shows the best fitting curve
      of DMRG data, corresponding to the value of $U$ which minimizes $\sigma$.
      We analyzed chains of $L=128$ sites,
      and evaluated correlations starting from the central site.}
    \label{fig:Phot-BH}
\end{figure}

A proper quantitative comparison between photon number correlations
in the coupled cavity system and boson number correlations in the BH
has been obtained by numerically solving the BH Bogoliubov equations for a
system with open boundaries, and it is shown in Fig.~\ref{fig:Phot-BH}.
The hopping is chosen to be equal in the two models ($t=J$);
the polariton/boson density is $\rho=1$. DMRG data are fitted 
by varying the on-site repulsion $U$ in the BH and by minimizing
the squared differences between the two curves:
$\sigma^2 (U) = \sum_r \vert C_r^{\rm ph} - C_r^{\rm BH} \vert^2$.
Of course, if $t/\beta$ is too small the Bogoliubov approximation fails,
and we do not find an appropriate minimum.
%
%%%%%%%%%%%%%%%%%%%%%%%%%%%%%
%%%%%%%%%%%%%%%%%%%%%%%%%%%%%
%
We remark that correlations are always negative, thus indicating
photon antibunching, as it is revealed by on-site number distributions
with negative values of the Mandel parameter~\cite{jcreview93}.

{\em Conclusions - } 
In this paper we studied photon and polariton fluctuations in coupled cavities. 
By a suitable scaling of the fluctuation detected over a region of the sample,
we showed that it is possible to extract the critical properties of the superfluid 
to Mott insulator phase transition. The analysis presented here for one-dimensional
systems can be extended to higher dimensional lattices. We remark that we confined
our study to the equilibrium case; an interesting point that remains to be 
addressed is to consider the effects of photon and cavity losses, which may lead 
to new non-equilibrium effects in the physics of such systems. We do however expect 
that the method proposed here will hold also in the non-equilibrium case. The 
reason is that our approach ultimately distinguishes the system being incompressible 
or not. While the general principle also holds for driven systems, the detailed
behaviour leading to the logarithmic scaling may be modified.
As a consequence, the most appropriate scaling ansatz for the non-equilibrium cases
has to be checked.

\acknowledgments

This work was supported by EU (EUROSQIP). We used the DMRG code of 
the ``Powder with Power'' project (\texttt{http://www.qti.sns.it/}).
We thank M. B. Plenio and I. Carusotto for interesting discussions.

\end{document}